\documentclass[10pt]{IEEEtran}
\IEEEoverridecommandlockouts
\usepackage{multirow}
\usepackage{makecell}
\usepackage{hyperref}

\usepackage{pdfpages}
\usepackage{textcomp}
\usepackage{epsfig,latexsym}
\usepackage{float}
\usepackage{indentfirst}
\usepackage{amsmath}
\usepackage{amssymb}
\usepackage{xcolor}
\usepackage{booktabs}

\usepackage{times}
\usepackage{subfigure}
\usepackage{psfrag}
\usepackage{cite}
\usepackage[ruled,longend,linesnumbered]{algorithm2e}
\usepackage{flushend}
\usepackage{epstopdf}
\usepackage{fancyhdr}
\usepackage{stfloats}
\usepackage{color}
\usepackage[noend]{algpseudocode}
\newtheorem{remark}{Remark}
\usepackage{setspace}
\usepackage{setspace}

\usepackage{geometry}
\begin{document}
\pagestyle{empty}

\title{Open Set RF Fingerprinting Identification: A Joint Prediction and Siamese Comparison Framework }
\author{\normalsize{\IEEEauthorblockN{Donghong Cai\IEEEauthorrefmark{1}, Jiahao Shan\IEEEauthorrefmark{1}, Ning Gao\IEEEauthorrefmark{2}, Bingtao He\IEEEauthorrefmark{3}, Yingyang Chen\IEEEauthorrefmark{1}, Shi Jin\IEEEauthorrefmark{4}, and Pingzhi Fan\IEEEauthorrefmark{5}}}

\IEEEauthorblockA{\IEEEauthorrefmark{1}\normalsize{College of Cyber Security and College of Information Science and Technology, Jinan University,
 Guangzhou  510632, China.}}

\IEEEauthorblockA{\IEEEauthorrefmark{2}The School of Cyber Science
and Engineering, Southeast University, Nanjing 210096, China.}

 \IEEEauthorblockA{\IEEEauthorrefmark{3}The State Key Laboratory of Integrated Services Networks, Xidian University, Xi'an 710071, China.}

\IEEEauthorblockA{\IEEEauthorrefmark{4}National Mobile Communications Research Laboratory, Southeast University, Nanjing 210096, China.}

\IEEEauthorblockA{\IEEEauthorrefmark{5}Institute of Mobile Communications, Southwest Jiaotong University,
 Chengdu 611756, China.}

 \IEEEauthorblockA{\normalsize{Email: dhcai@jnu.edu.cn; shanjh@stu2022.jnu.edu.cn; ninggao@seu.edu.cn; bthe@xidian.edu.cn; chenyy@jnu.edu.cn; jinshi@seu.edu.cn; pzfan@swjtu.edu.cn.}}
\vspace{-1.5em}
}

\maketitle
\thispagestyle{empty}
\begin{abstract}
Radio Frequency Fingerprinting Identification (RFFI) is a lightweight physical layer identity authentication technique. It identifies the radio-frequency device by analyzing the signal feature differences caused by the inevitable minor hardware impairments. However, existing RFFI methods based on closed-set recognition struggle to detect unknown unauthorized devices in open environments. Moreover, the feature interference among legitimate devices can further compromise identification accuracy. In this paper, we propose a joint radio frequency fingerprint prediction and siamese comparison (JRFFP-SC) framework for open set recognition. Specifically, we first employ a radio frequency fingerprint prediction network to predict the most probable category result. Then a detailed comparison among the test sample's features with registered samples is performed in a siamese network. The proposed JRFFP-SC framework eliminates inter-class interference and effectively addresses the challenges associated with open set identification. The simulation results show that our proposed JRFFP-SC framework can achieve excellent rogue device detection and generalization capability for classifying devices.
\end{abstract}
\vspace{-0.2em}
\begin{IEEEkeywords}
Radio frequency fingerprinting identification, open set,  physical layer authentication, siamese network.
\end{IEEEkeywords}

\vspace{-0.7em}
\section{Introduction}
\IEEEPARstart{W}{ith} the wide application of the Internet of Things (IoT), massive devices connects to the communication networks. Compared to wired networks, the openness of wireless networks allows the potential devices to access the network flexibly, leading to unauthorized access and potential security threats. Especially, many IoT devices are unable to effectively employ encryption algorithms due to the transmission delay requirement and limited computation resource \cite{neshenko2019demystifying}.
Physical Layer Authentication (PLA) leverages the dynamics of physical layer attributes to address authentication-based attack challenges and enhance wireless security\cite{wang2016physical}. By utilizing the random characteristics of wireless channels and the physical layer hardware features of devices, PLA can fundamentally improve the security of identity recognition with lower overhead, such as Channel State Information (CSI)\cite{xiao2008using,fang2018learning}, Received Signal Strength Indicator (RSSI)\cite{10682524}, and Radio Frequency (RF) fingerprint\cite{roy2019rfal,xue2022radio}. RF fingerprinting leverages the intrinsic hardware imperfections present in RF components that arise during the manufacturing process. The hardware imperfections of these components have slight deviations from their standard values, which will not affect normal communication functions\cite{shen2022towards}. This renders RF components challenging to duplicate, complicates the forgery of their unique identifiers, and allows specific targeting. The receiver identifies the unique identity of the device by extracting the RF fingerprint without the need for additional network protocols or encryption mechanisms. 

Recently, deep learning has empowered radio frequency fingerprint identification (RFFI) due to its outstanding feature extraction capabilities \cite{shen2023deep}. However, most of the existing works focus on the closed-set to train and test deep learning models for identifying RF fingerprint, which is not conducive to detecting potential unauthorized identities in wireless networks. Specifically, it assumes that the identities of the sources during the training process are known, and the signals during the testing process come from unseen data packets with known identities. Even if the known dataset covers a large number of identities, an unseen source enters the communication network, it will be misidentified as a known identity. Additionally, the softmax layer is often considered as the decision layer for classification\cite{9155259}. However, once the number of classes is determined in softmax, it becomes non-expandable\cite{shen2022towards}, resulting in the trained model lacking the ability to perceive unseen identity data. Therefore, RFFI necessitates more open set experimental conditions\cite{xie2021generalizable,hanna2020open}.

\begin{figure}[tp]
	\centering
    \subfigure[Ideal situation]{
    \label{Fig1a}
	\includegraphics[width=3.0cm,height = 1.6cm]{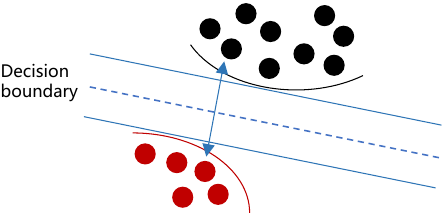}}
	\subfigure[Interference between legitimate classes]{
    \label{Fig1b}
    \includegraphics[width=5.0cm,height = 2.6cm]{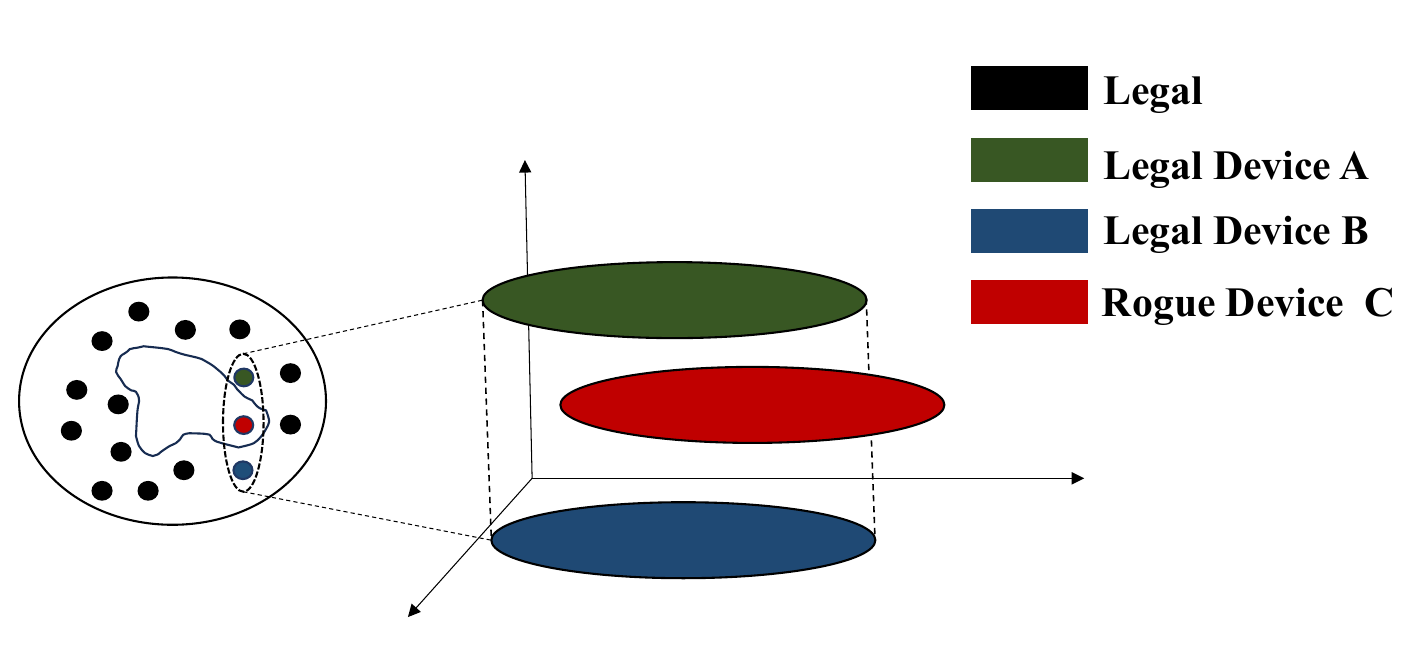}}
    \caption{Illustration of legitimate device feature overlap.}\label{Fig1}
\end{figure}

For rogue detection of open set RF fingerprint, it is practical to analyze the RF characteristics of a known source and compare it to observed samples\cite{shen2022towards}. A match is deemed unsuccessful if the observed deviation surpasses the predetermined decision boundary shown in Fig. \ref{Fig1a}. Thus more knowledge of the characteristics associated with all legitimate categories is required to filter out data packets that do not correspond to the categories. However, the presence of shared characteristics among legitimate categories may hinder the precise identification of unauthorized entities. As shown in Fig. \ref{Fig1b}, there are legitimate devices A and B, their RF signals are overlapped in the feature space. In fact, the feature vector of device A's RF signal can be transformed to approximate the feature vector of device B in a certain feature dimension. If a rogue device C has an RF signal feature vector that falls between those of devices A and B in the feature space, detection algorithms relying on all legitimate devices will be affected by the overlapping characteristics of A and B. This overlap makes it challenging to accurately distinguish rogue devices, as the shared features among legitimate devices create interference and reduce identification accuracy.

In this paper, we consider the open set RFFI, and a joint radio frequency fingerprint prediction and siamese comparison (JRFFP-SC) framework is proposed, which seeks to identify rogue identities by analyzing the differences between observed samples and their corresponding inferred categories. Specifically, we designed a radio frequency fingerprint prediction network (RFFP) using VGG11\cite{simonyan2014very} and fully connected layers, which is capable of extracting RF fingerprints and has a high accuracy in terms of fingerprint classification. This means that among all possible identities, the observed sample identity is more similar to the inferred identity. However, it does not mean that the sample comes from this identity, because there are also rogue devices present. Therefore, by conducting a comparative analysis between the observed sample and the average sample of the corresponding identity recorded in the database through siamese comparison, it is possible to ascertain whether the observed sample originates from unauthorized RF devices. The experiment results show that the proposed JRFFP-SC framework achieves a high accuracy rate for legitimate identities while maintaining a high detection rate for rogue identities.

\vspace{-0.4em}
\section{System Model and Data Preprocessing}
\subsection{System Model}
We consider the signal source from LoRa devices, which uses chirp spread spectrum (CSS) as the modulation technique. Without loss of generality, it is assumed that the header of each signal packet contains eight repeated preambles, represented as
\begin{align}
O(t)=Me^{j{\pi}t(-BW+(BW){R_s}t)},
\end{align}
where $t \in [0,T]$, $M$ represents the amplitude, $BW$ represents the bandwidth. In addition, $R_s$ represents the symbol rate of LoRa modulation, expressed as
\begin{align}
R_s=\frac{BW}{2^{SF}},
\end{align}
where $SF$ stands for spreading factor. The baseband signal $x(t)$, composed of eight preambles, is transmitted to the receiver via wireless communication through a series of amplifiers, such as power amplifiers, low-noise amplifiers, and oscillators, etc. However, these hardware components may have hardware impairment, which results in minor disruptions to the transmission signal. The degree of hardware impairment differs across various devices, consequently leading to varying degrees of disturbance. Consequently, device identity recognition can be achieved by
\begin{align}
y(t)=h(t)*\mathcal{G}(x(t))+n(t),
\end{align}
where $h(t)$ is the impulse response of the wireless channel, $\mathcal{G}(\cdot)$ represents the disruption function resulting from hardware impairment, $n(t)$ denotes additive Gaussian white noise, and * denotes the convolution operation.

\subsection{Data Preprocessing}
In order to improve the quality of the training data, it is essential for the signals captured by the receiver to undergo a series of processing steps, including synchronization, extraction of the preamble code, compensation for carrier frequency offset (CFO), and normalization. Subsequently, the time-frequency characteristics of the received signal $y(t)$ are derived through segmentation and the application of the short-time Fourier transform (STFT) to each segment. Specifically, a window function $w(t)$ is employed to extract the local features of the signal around time $t$, thereby obtaining the STFT representation $Y(t,f)$ at the corresponding frequency $f$ for each time $t$, i.e,
\begin{align}
Y(t,f)=STFT(y(t))=\sum_{n=0}^{N}y(n)w(t-n)e^{-j2{\pi}fn},
\end{align}
where $n$ represents each sampling point of the discrete-time signal. The STFT matrix $\mathbf{S}$ can be represented as
\begin{align}
\mathbf{S}=\{Y(t,f)|t \in T, f \in F\}.
\end{align}
By calculating the magnitude of the short-time Fourier transform matrix $\mathbf{S}$, we obtain the frequency spectrum of the signal, denoted as
\begin{align}
\tilde{\mathbf{S}}=10\log_{10}{|\mathbf{S}|^2}.
\end{align}
The collected spectrograms are classified and organized into the training dataset $\mathcal{D}_{train}$ according to the identity source, represented as
\begin{subequations}
\begin{align}
&\mathcal{D}_{train}=\{d_i\}_{i=1}^I,\\
&d_i= \{(\tilde{S}_n,l_n)\}_{n=1}^N,
\end{align}
\end{subequations}
where $I$ is the identity type of the legitimate RF device, $N$ denotes the amount of data corresponding to each category, $l_n$ represents the label for each observed sample.
\section{Proposed JRFFP-SC Framework}
\begin{figure}[!h]
	\centering
	\vspace*{0.5em}
	\includegraphics[width=1.0\linewidth]{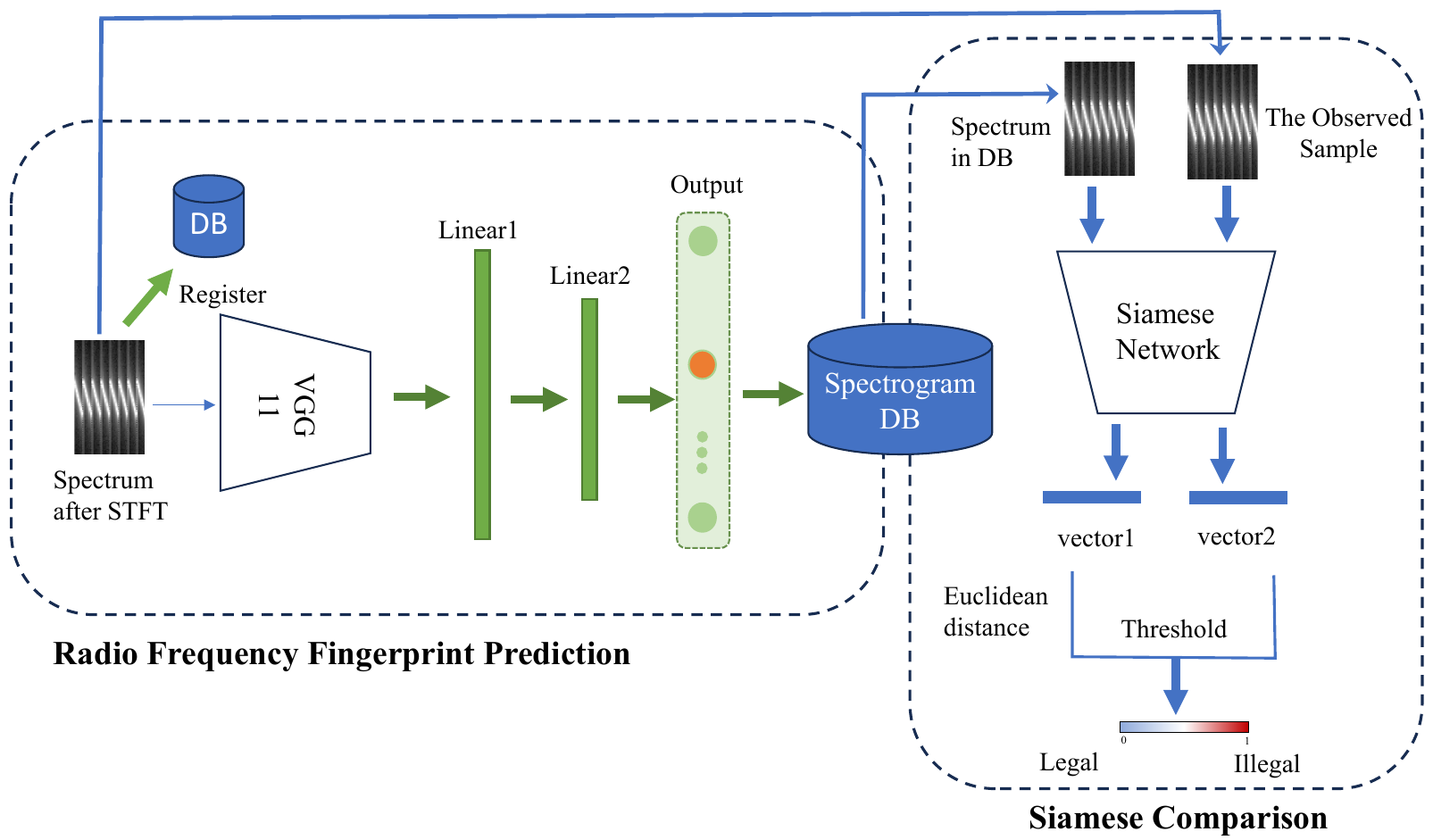}
	\vspace*{-0.5em}
	\caption{Proposed JRFFP-SC framework.}\label{fig:3}
\end{figure}

In this section, we propose a novel JRFFP-SC framework for RF fingerprint extraction and identity recognition. As shown in Fig. \ref{fig:3}, the proposed JRFFP-SC for open set RFFI contains a RFFP network and a siamese (SIA) network\cite{bromley1993signature}.

\subsection{Network Design of Proposed JRFFP-SC Framework}

\textbf{RFFP Network:} RFFP will extract essential characteristics from the spectrogram to create a unique RF fingerprint using VGG11 and predicting the RF identity through a pre-recognition network.

\begin{figure}[tp]
	\centering
    \subfigure[RFFP network.]{
    \label{Fig4a}
	\includegraphics[width=4.0cm,height = 2.8cm]{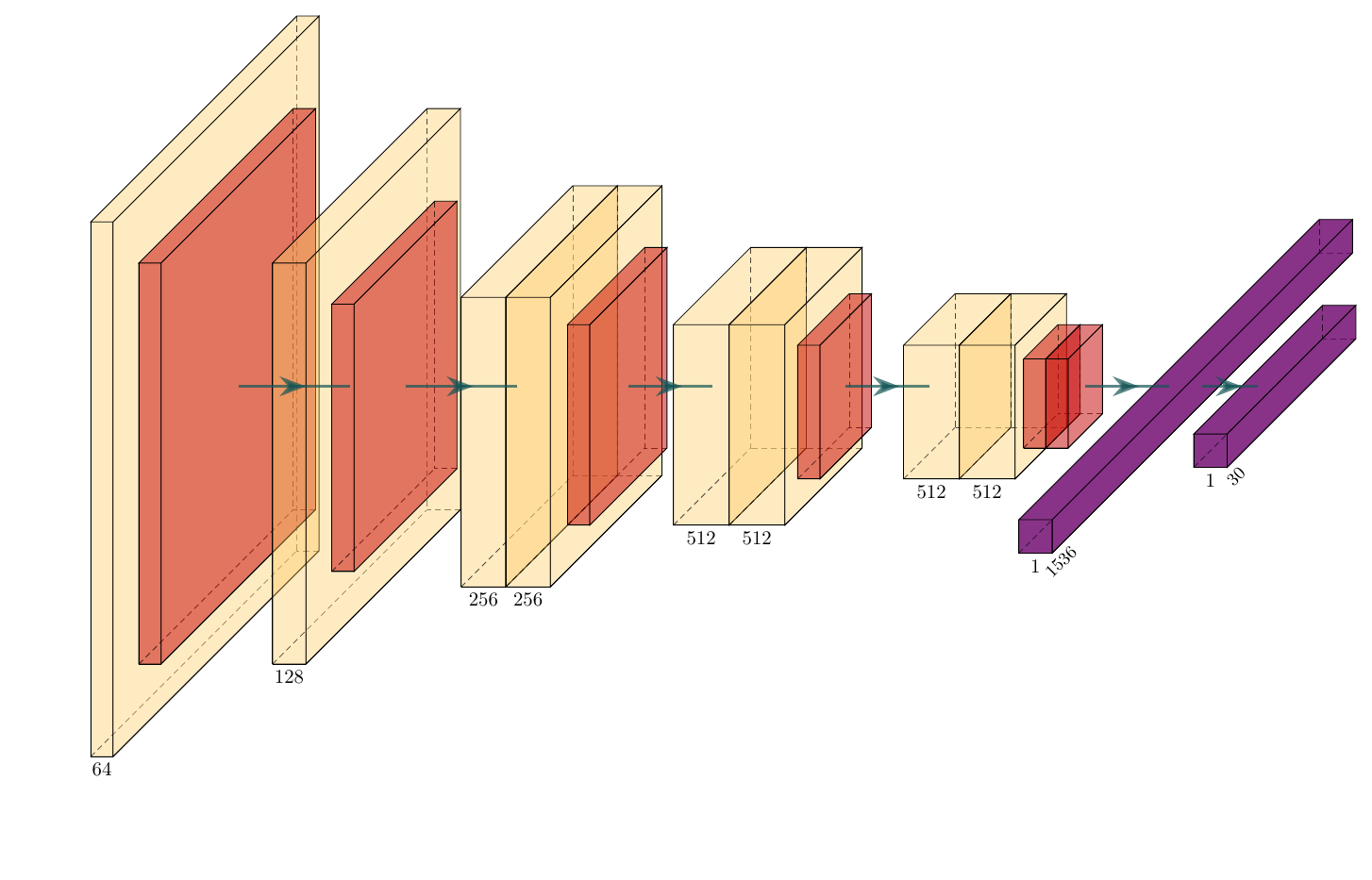}}
	\subfigure[Siamese network.]{
    \label{Fig4b}
    \includegraphics[width=4.0cm,height = 2.8cm]{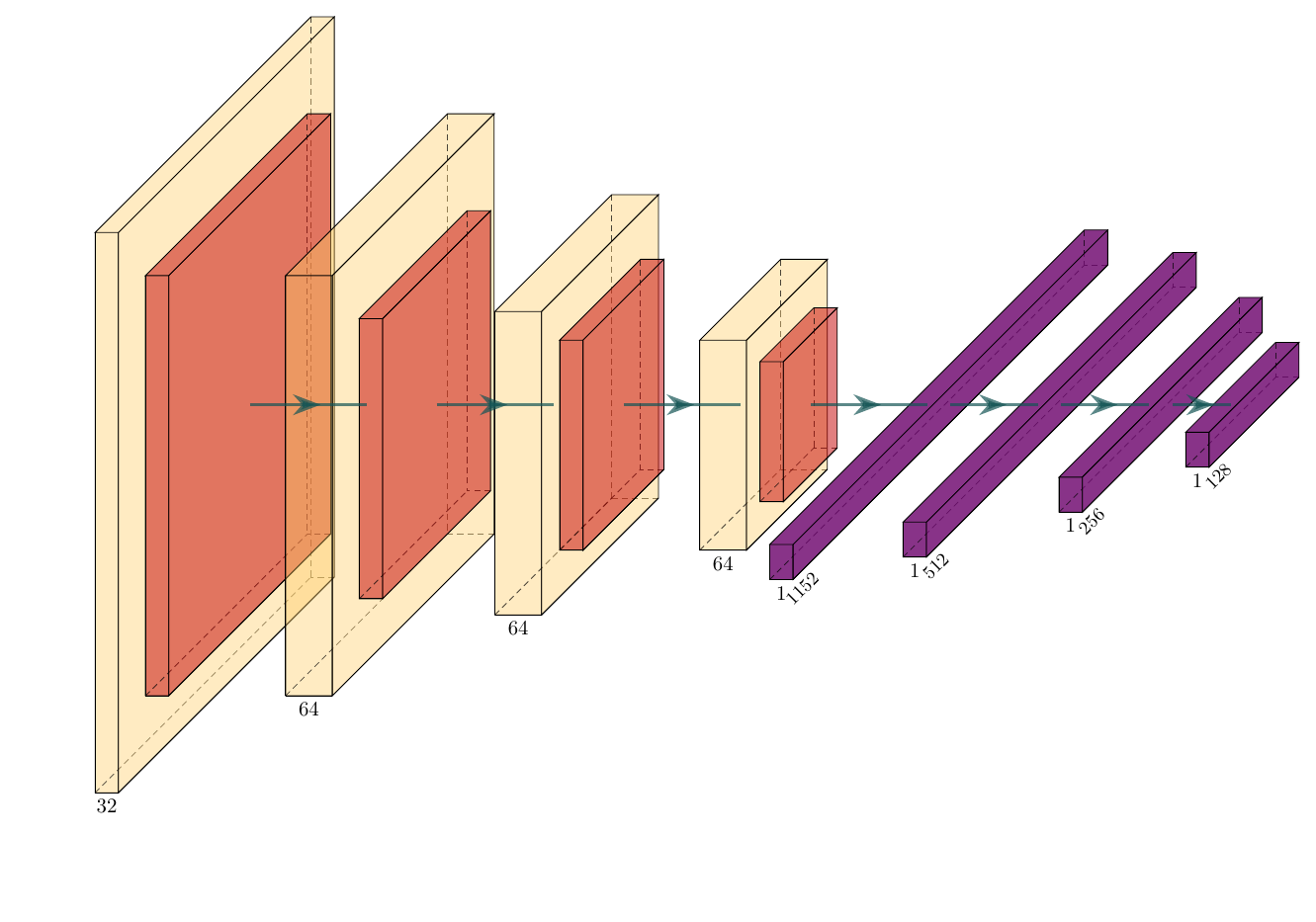}}
    \caption{The separate network structures of proposed JRFFP-SC framework.}\label{Fig4}
\end{figure}

As shown in Fig. \ref{Fig4a}, the VGG11 of proposed JRFFP-SC framework analyzes the features of spectrogram at multiple scales through convolution operators. In particular, it employs 8 stacked convolutional layers with $3\times3$ convolutional kernels. The pooling layers are $2\times2$ max pooling and $1\times1$ average pooling. After feature extraction by VGG11, the original spectrogram is transformed into a $512\times3$ RF fingerprint. This fingerprint is then processed through two fully connected layers, and the final identity probabilities are computed using a softmax function.

The optimization objective of network $\mathcal{F}_{RFFP}$ on the training set $D_{train}$ is defined as
\begin{equation}\label{opt1}
\Omega^{\mathrm{opt}} \stackrel{\mathcal{L}_\Omega}{\gets} \mathcal{F}_{RFFP}(\mathfrak{D}_{train}, \Omega),
\end{equation}
where $\mathcal{L}$ represents the loss function, $\Omega$ is the network's parameters.

\textbf{SIA Network:} SIA is a further assessment of the recognition results from the RFFP. It compares the average spectrograms of registered samples in the database to determine whether the observed sample belongs to the predicted category. In particular, the SIA can eliminate feature interference among legitimate devices, and achieving effective identification of rogue devices. SIA consists of four layers of $3\times3$ convolutional layers and three fully connected layers as shown in Fig. \ref{Fig4b}.

The optimization objective of network $\mathcal{F}_{SIA}$ on the training set $\mathfrak{D}^{enroll}$ is defined as
\begin{equation}
\Theta^{\mathrm{opt}} \stackrel{\mathcal{L}_\Theta}{\gets} \mathcal{F}_{SIA}((\mathfrak{D}_{train}, \mathfrak{D}^{enroll}), \Theta),
\end{equation}
where $\mathcal{L}$ represents the loss function, $\Theta$ is the network's parameters.

\subsection{Network Training and Inference}
\begin{figure}[tp]
	\centering
	\vspace*{0.5em}
	\includegraphics[width=1.0\linewidth]{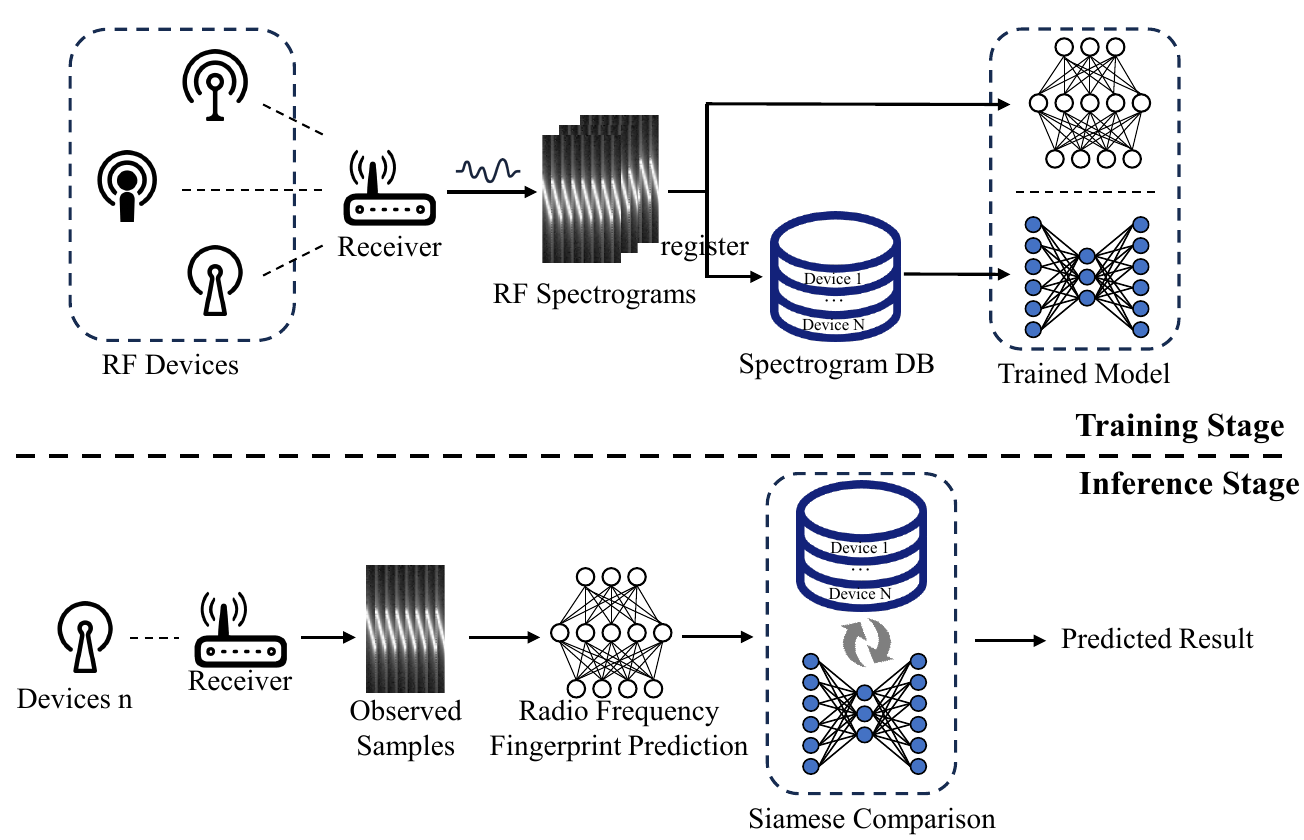}
	\vspace*{-0.5em}
	\caption{Training and inference illustrations of proposed JRFFP-SC framework.}\label{fig:2}
\end{figure}

The system is divided into two main phases: the training phase and the inference phase, as shown in the Fig. \ref{fig:2}.
\subsubsection{Training}
The proposed JRFFP-SC framework collects labeled RF signal packets from multiple RF devices. After these packets are captured by the receiver, they are converted into spectrograms. The spectrograms are divided according to identity sources, and the average spectrogram for each identity is obtained by
\begin{align}
d_i^{enroll}=\frac{1}{N}\sum_{n=1}^{N}d_{i,n},
\end{align}
where $i$ represents the identity type of the legitimate RF device, and $N$ indicates the data quantity per category.

The average spectrogram will be stored in the spectrogram database to represent the registration of the identity of legitimate devices, denoted as $\mathcal{D}^{enroll}=\{d_i^{enroll}\}_{i=1}^I$.

Subsequently, the RFFP will extract the RF fingerprint from the spectrogram and obtain the identity probability density vector, which can be described as
\begin{align}
\hat{\rho}=\mathcal{F}_{RFFP}(d_p,\Omega),
\end{align}
where $d_p$ represents the spectrogram of identity $p$, and $\hat{\rho}$ denotes the probability distribution vector obtained through RFFP. Furthermore, the maximum component $\hat{p}$ of the $\hat{\rho}$ vector reveals the prediction identity of the observed sample can be described by
\begin{align}
\hat{p} \gets
\begin{cases}
\mathrm{arg}\max(\hat{\rho}), & \text{with probability } 0.5, \\
random(I), & \text{with probability } 0.5.
\end{cases}
\end{align}

It is important to emphasize that during the training process, random identities are generated with 50\% probability to simulate rogue device samples. The loss function for backpropagation is the cross-entropy loss, given by
\begin{align}
\mathcal{L}_{ce}=-\sum_{i=1}^{I}\rho_{i}\log_{\hat{\rho}_i}.
\end{align}

Furthermore, we optimize the model parameters by minimizing this loss by
\begin{align}
\Omega \gets \Omega - \zeta\nabla_{\Omega}(\mathcal{L}_{ce}),
\end{align}
where $\zeta$ denotes learning rate and $\nabla$ means gradient descent algorithm.
SIA can represent the difference in feature vectors between the observed sample and the $\hat{p}$ samples in the registered database as
\begin{align}
(v_{\hat{p}},v_p)=\mathcal{F}_{SIA}((d_{\hat{p}}^{enroll}, d_p), \Theta),
\end{align}
where $v_{\hat{p}}$ and $v_p$ are the low-dimensional feature vectors derived from the registered spectrogram corresponding to identity $\hat{p}$ and the spectrogram corresponding to observed samples for identity $p$, respectively. The difference between two feature vectors can be compared using the following contrastive loss, i.e.,
\begin{align}
\mathcal{L}_{con}=\frac{1}{I}\sum_{i=1}^I[(1-a_i)*D^2+a_i*max(0,m-D^2)],
\end{align}
where $D=\Vert v_{\hat{p}}-v_p \Vert_2$, $a_i= p \oplus \hat{p}$ represents legitimacy label, and $m$ is a boundary hyperparameter used to distinguish dissimilar sample pairs. The parameter optimization of SIA can be expressed as
\begin{align}
\Theta \gets \Theta - \zeta\nabla_{\Theta}(\mathcal{L}_{con}).
\end{align}
The detail training of the proposed JRFFP-SC framework is shown in Algorithm \ref{alg:1}.

\begin{remark}
SIA employs RF spectrogram to train and obtain more comprehensive information about signals, rather than relying on the RF fingerprint obtained from RFFP extractor.
\end{remark}

\begin{algorithm}[tp]\small
    \caption{Training of proposed JRFFP-SC}
    \label{alg:1}
    \KwIn{Dataset $\{d_i\}_{i=1}^I \in \mathfrak{D}_{train}$, initial parameters $\Omega, \Theta$, learning rate $\zeta$.}
    \KwOut{The trained parameters $\Omega^{opt}, \Theta^{opt}$.}
        \For{each legitimate identity $i \in I$}{
            // Registered average spectrograms                           \\
            $\mathfrak{D}^{enroll}[i] \gets mean(\mathfrak{D}_{train}[i])$; \\
        }
        \Repeat{Convergence of parameters $\Omega$, $\Theta$}{
            \For{each $d_p \in \mathfrak{D}_{train}$}{
                // Train RFFP Network   \\
                $\hat{\rho}=\mathcal{F}_{RFFP}(d_p,\Omega);$      \\
                $\mathcal{L}_{ce}=-\sum_{i=1}^{I}\rho_{i}\log_{\hat{\rho}_i};$ \\
                $\Omega \gets \Omega - \zeta\nabla_{\Omega}(\mathcal{L}_{ce});$  \\
                $ $ \\
                // Train SIA Network  \\
                $\hat{p} \gets \begin{cases} \mathrm{arg}\max(\hat{\rho}), & \text{with probability } 0.5 \\random(I), & \text{with probability } 0.5 \end{cases};$  \\
                $d_{\hat{p}}^{enroll} \gets \mathfrak{D}^{enroll}[\hat{p}];$  \\
                $(v_{\hat{p}},v_p)=\mathcal{F}_{SIA}((d_{\hat{p}}^{enroll}, d_p), \Theta);$ \\
                $\mathcal{L}_{con}=\frac{1}{I}\sum_{i=1}^I[(1-a_i)*D^2+a_i*max(0,m-D^2)];$ \\
                $\Theta \gets \Theta - \zeta\nabla_{\Theta}(\mathcal{L}_{con});$  \\
            }
        }
\end{algorithm}

\begin{algorithm}[tp]\small
    \caption{Inference based on proposed JRFFP-SC}
    \label{alg:2}
    \KwIn{Dataset $\{d_i\}_{i=1}^I \in \mathfrak{D}_{test}, \{d_i^{enroll}\}_{i=1}^I \in \mathfrak{D}^{enroll}$, trained parameters $\Omega^{opt}, \Theta^{opt}$, threshold $\lambda$}
    \KwOut{Decision}

    \For{each $d_p \in \mathfrak{D}_{test}$}{
        $\hat{\rho}=\mathcal{F}_{RFFP}(d_p,\Omega^{opt});$ \\
        $\hat{p} \gets argmax(\hat{\rho});$         \\
        $d_{\hat{p}}^{enroll} \gets \mathfrak{D}^{enroll}[\hat{p}];$  \\
        $(v_{\hat{p}},v_p)=\mathcal{F}_{SIA}((d_{\hat{p}}^{enroll}, d_p), \Theta^{opt});$ \\
        $D = \Vert v_{\hat{p}} - v_p \Vert_2$   \\
        \If{$\lambda < D$}{
          Decision = identity is rogue device;
        }
        \Else{
          Decision = identity is p;
        }
    }
\end{algorithm}
\subsubsection{Inference}
The RF signals in the test set are open set, with some unseen categories acting as rogue classes. These signals are captured by the receiver and converted into RF spectrograms.

The trained RFFP is used to extract RF fingerprint features from the spectrograms and identify the most likely identity of the fingerprint, i.e., Algorithm \ref{alg:2} line 2 to 3. Lines 4 to 6 of Algorithm \ref{alg:2} are crucial to the SIA for eliminating inter-class interference. In contrast to the approach that involves comparing the observed sample with all potential identity samples, this method concentrates on a particular possible identity to facilitate a comparison between the two categories. Specifically, based on the trained SIA network, the similarity error $D$ between the registered samples of that identity in the spectrogram database and the observed samples is compared. A reasonable threshold $\lambda$ is selected; if the error exceeds the threshold, the observed sample is determined to come from a rogue device. Otherwise, the prediction provided by the RFFP is accepted as the definitive result, i.e.,
\begin{equation}
\mathrm{Decision}=\left\{
\begin{aligned}
&\mathrm{Rogue} \ \mathrm{Device},& \!\!\!\!\mathrm{for} \ \lambda < D, \\
&\mathrm{Legitimate} \ \mathrm{Device},& \mathrm{for} \ \lambda >= D. \\
\end{aligned}
\right.
\end{equation}

\vspace{-0.5em}
\section{Simulation Result}
\label{Sec5}

In this section, we first introduce the experimental dataset and environment. Then, the metrics used to measure the performance of the framework and the experimental results are presented.

Dataset\footnote{\url{https://github.com/gxhen/LoRa_RFFI/tree/main/Openset_RFFI_TIFS}} \cite{shen2022towards,shen2023deep} contains RF signals collected from 45 LoRa devices. The LoRa devices is Lopy4 equipped with the SX1276 chip, and the data was collected at a fixed indoor Line of Sight (LOS) position. Among them, devices numbered 1 to 30 server as legitimate RF devices. In particular, to enhance the network model's generalization ability, data augmentation was applied to the training set, adding multipath fading and Doppler shift. Devices numbered 31 to 45 are employed as unauthorized devices to assess the identification efficacy of the SIA. The test set comprises categories that the trained model has encountered as well as previously unobserved rogue categories, thereby identifying it as an open set. Furthermore, rogue devices constitute approximately 29.41\% of the test set. The scalable channel robust radio frequency fingerprint identification (SC-RFFI) proposed in \cite{shen2022towards} is considered as a baseline that demonstrates excellent scalability and impressive accuracy in identity recognition. In addition, our experiments are conducted on an AMD-R5 5800H CPU, Nvidia RTX 3050Ti, and Windows 11 OS.
To evaluate the proposed JRFFP-SC, the following performance are considered:
\begin{itemize}
\item {\bf{Classification}}: Accuracy is the proportion of correctly classified samples. WThe confusion matrix shows the consistency between the actual and predicted values. Each row of the matrix represents the number of 400 samples. A higher concentration of values along the diagonal indicates an increased accuracy in classification.
\item {\bf{Rouge Device Detection}}: Receiver operating characteristic curve (ROC) illustrates the relationship between the false positive rate (FPR) and the true positive rate (TPR) across various threshold settings. Area under the curve (AUC) and equal error rate (EER) are two key indicators. A higher AUC signifies better performance, while a lower EER (where FPR = 1 - TPR) indicates enhanced accuracy.
\end{itemize}

\begin{figure}[tp]
	\centering
    \subfigure[The classification accuracy of SC-RFFI is $95.33\%$.]{
    \label{Fig5a}
	\includegraphics[width=4.2cm, height=3.5cm]{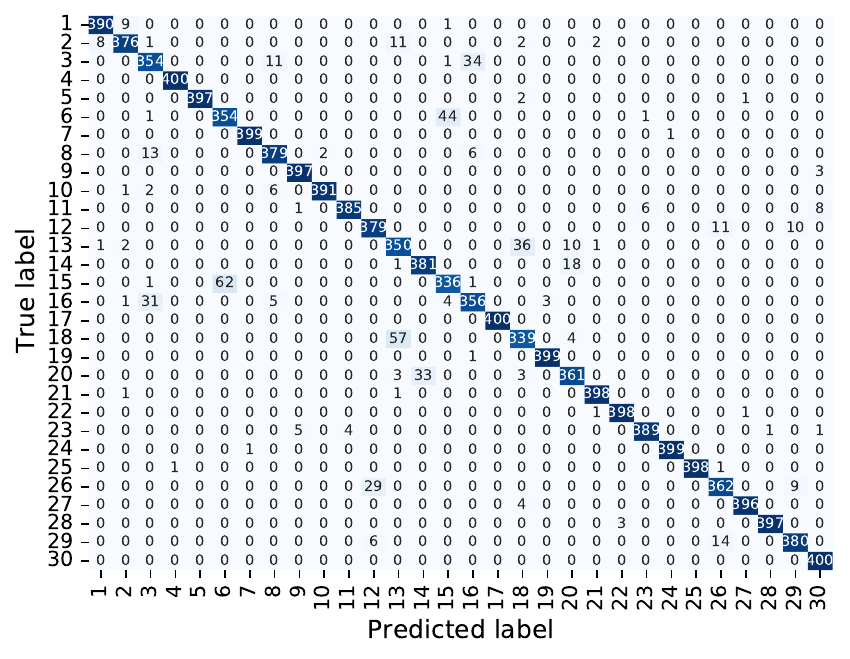}}
	\subfigure[The classification accuracy of JRFFP-SC is $98.47\%$.]{
    \label{Fig5b}
    \includegraphics[width=4.2cm, height=3.5cm]{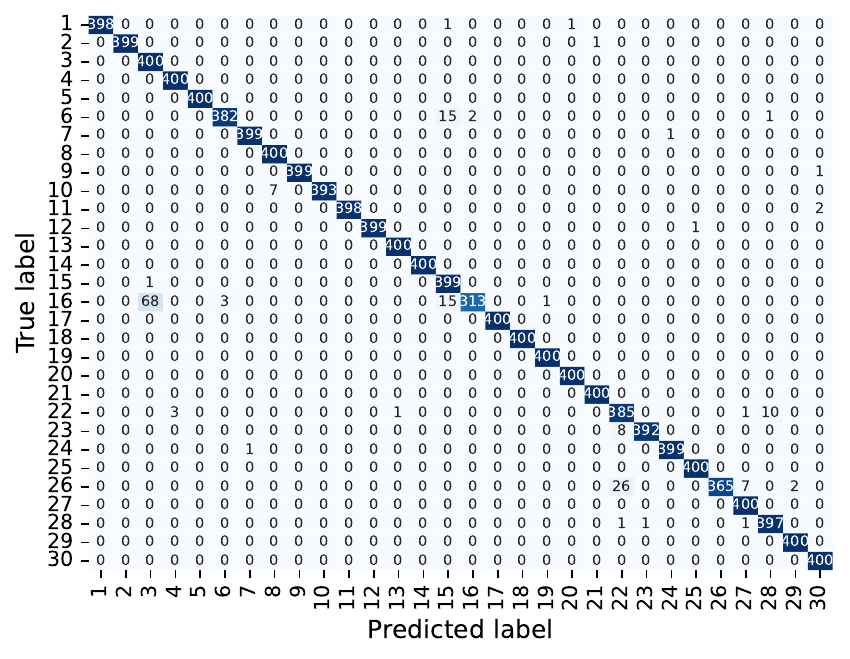}}
    \caption{Classification results for 30 legal categories.}\label{Fig5}
\end{figure}

As shown in Fig. \ref{Fig5}, we assess the efficacy of legitimate identity recognition using the device test set numbered 1-30 with the confusion matrix. The SC-RFFI method attains an accuracy rate of 95.33\%. The JRFFP-SC achieves an accuracy rate of $98.47\%$. This indicates that the RFFP has higher level of accuracy, thereby offering a more dependable foundation for siamese comparison. Compared to the baseline, our proposed JRFFP-SC framework performs better in the confusion matrix. The actual and predicted labels exhibit a more consistent alignment along the diagonal, while the instances of misclassification off the diagonal are reduced. This indicates that the RFFP is capable of accurately classifying samples.

\begin{figure}[tp]
	\centering
	\vspace*{0.5em}
	\includegraphics[width=0.8\linewidth]{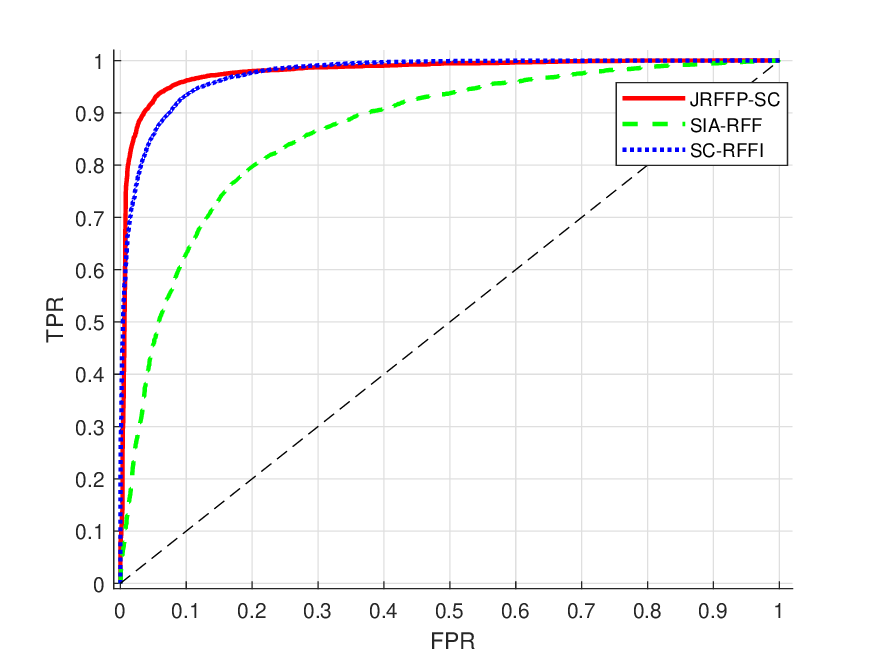}
	\vspace*{-0.5em}
	\caption{ROC curve for the identification of unauthorized devices. The AUC of JRFFP-SC is 0.979, and the EER is 0.061; the AUC of SC-RFFI is 0.968, and the EER is 0.091; the AUC of SIA-RFF is 0.914, and the EER is 0.147.
}\label{fig:6}
\end{figure}

Moreover, we evaluated three different methods: JRFFP-SC, SIA-RFF and SC-RFFI  on a device test set ranging from 1 to 45 in the unauthorized device identification experiment. Notably, SIA-RFF denotes the SIA that has been registered and trained utilizing RF fingerprints extracted through the RFFP. The characteristics of this approach are its lightweight nature and fast response speed. As shown in Fig. \ref{fig:6}, the JRFFP-SC method has the highest AUC (0.979) and the lowest EER (0.061), indicating that it performs best overall at different thresholds. Additionally, JRFFP-SC excels in balancing the false positive rate and the false negative rate.

However, the performance of SIA-RFF is lower than the other two methods. Despite its foundation in SIA, SIA-RFF exhibits a lower AUC due to the use of RF fingerprints for training, which may have removed some key features that are helpful for unauthorized device identification.
Table. \ref{tab:1} presents more detailed indicator information, including Accuracy, Precision, Recall, F1-Scores and Time (s/1000 records). The JRFFP-SC has the highest accuracy and recall, which makes it potentially more reliable and stable in practical applications. The SC-RFFI model demonstrates the highest precision and F1 score, suggesting its superiority in achieving a balance between accuracy and recognition rate. Although the performance of SIA-RFF needs improvement, it takes the least time to process every thousand data entries (0.0043s), which may be suitable for scenarios requiring quick responses, i.e., the industrial IoTs.

\begin{figure*}[tp]
	\centering
    \subfigure[The impact of SNR on the accuracy of legitimate device classification.]{
    \label{Fig7a}
	\includegraphics[width=5.8cm,height = 4.5cm]{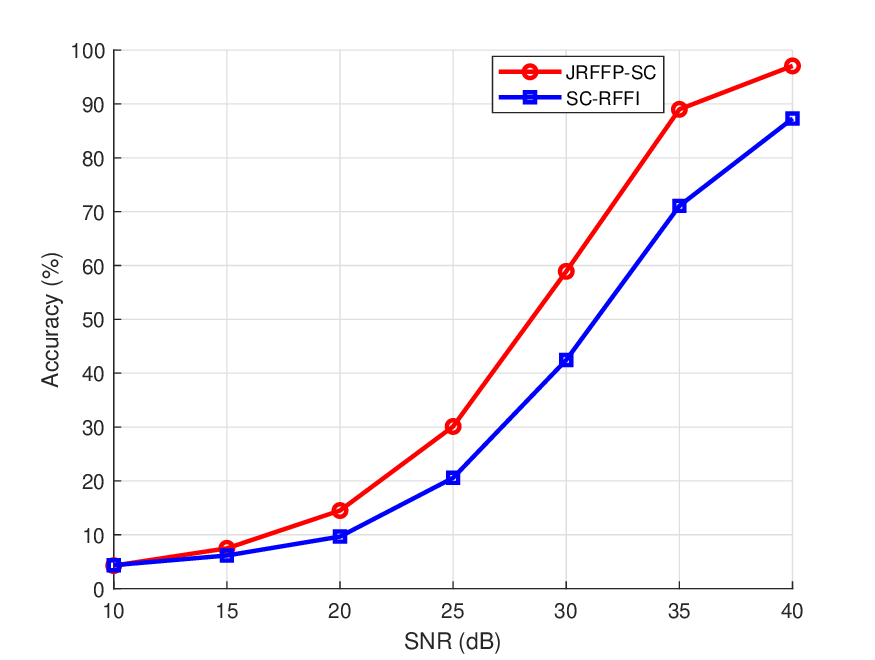}}
	\subfigure[The impact of SNR on the accuracy of rogue device recognition.]{
    \label{Fig7b}
    \includegraphics[width=5.8cm,height = 4.5cm]{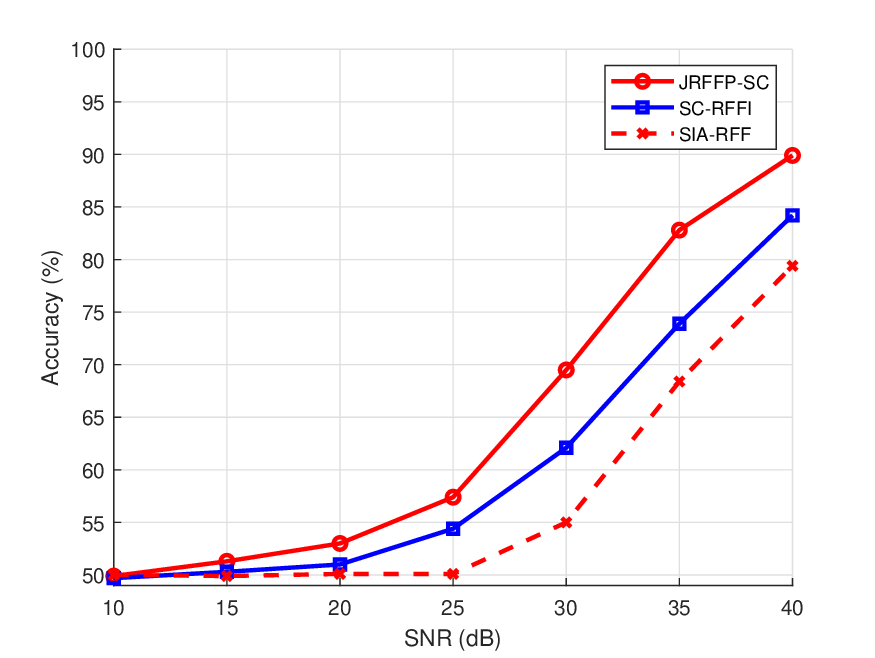}}
	\subfigure[The impact of SNR on the AUC curve for rogue device detection.]{
    \label{Fig7c}
    \includegraphics[width=5.8cm,height = 4.5cm]{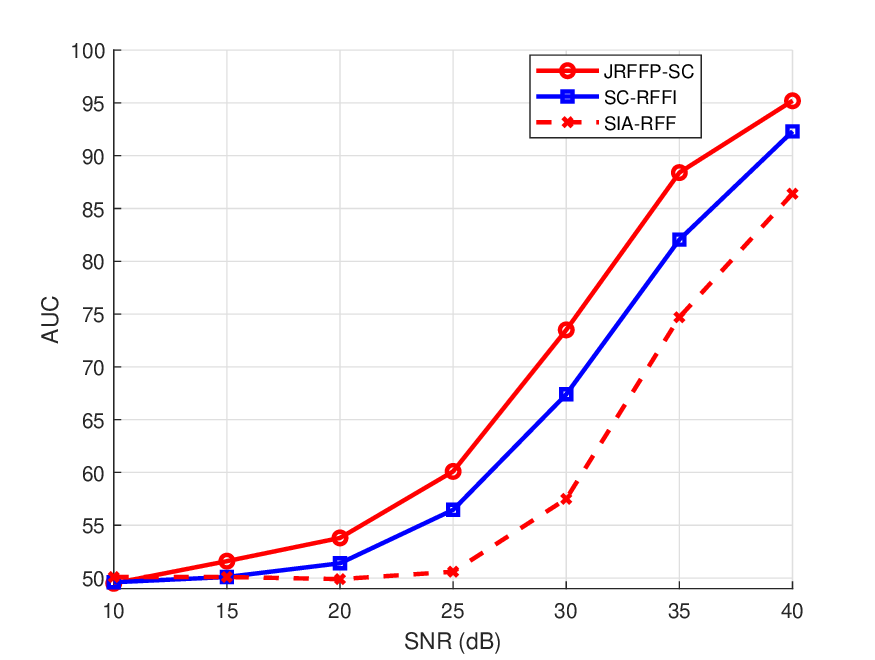}}
    \caption{The impact of SNR on legitimate device classification and open set device detection.}\label{Fig7}
\end{figure*}

\begin{table}
\centering
\caption{More detailed indicators for rogue device detection.}
\renewcommand\arraystretch{1.2}
\setlength{\tabcolsep}{1.2mm}{
\begin{tabular}{cccccc}
\toprule
Method \!\! & Accuracy & \!Precision & Recall & F1-Scores & \!Time\\
\hline
JRFFP-SC     & 0.939    & 0.864     & 0.939  & 0.900     & 1.4749                 \\
SC-RFFI\cite{shen2022towards} & 0.916    & 0.963     & 0.916  & 0.939     & 2.7305                 \\
SIA-RFF & 0.853    & 0.708     & 0.853  & 0.774     & 0.0043                 \\
\bottomrule
\end{tabular}
}\label{tab:1}
\end{table}

Since the experimental test set received data packets in a high signal-to-noise ratio (SNR) environment, it is considered noise-free data. To explore the impact of different SNRs on model performance, we added additive gaussian white noise to simulate different SNR conditions during the testing process. The experimental results are shown in the Fig. \ref{Fig7}. Our framework achieved higher classification accuracy and rogue recognition accuracy than the baseline model. This indicates that our proposed framework has good robustness to SNR variations. Additionally, the SIA using RF fingerprints as training input performed poorly and lost its capability in low SNR environments.

\section{Conclusion}
This article addresses the issues of identity recognition and interference between legitimate device categories in open set environments. We proposed a joint radio frequency fingerprint prediction and siamese comparison (JRFFP-SC) framework. Initially, a VGG11-based radio frequency fingerprint prediction network forecasts the most probable category of legitimate devices. Subsequently, employing the siamese network, it compares the feature similarity between test samples and registered samples to eliminate interference among legitimate devices and distinguish rogue devices. Experimental results show that the proposed JRFFP-SC can effectively identify rogue devices and legitimate devices in open set environments. The experiments also validate the robustness of JRFFP-SC under different SNRs.

\vspace{-0.5em}
\section*{Acknowledgment}
The work of D. Cai was supported by the Science and Technology Major Project of Tibetan Autonomous Region of China (No. XZ202201ZD0006G02), the Basic and Applied Basic Research Foundation of Guangdong province (No. 2024A1515012398) and National Science
Foundation of China (NSFC) (No.62001190). The work of N. Gao was supported by the NSFC (No.62371131), the program of Zhishan Young Scholar of Southeast University (No.2242024RCB0030) and the Fundamental Research Funds for the Central Universities (No.2242023K5003). The work of B. He was supported by the NSFC (No.62201421).
\vspace{-0.5em}
\bibliographystyle{IEEEtran}
\bibliography{IEEEfull,trans}

\end{document}